\documentclass[journal]{IEEEtran}

\pdfoutput=1
\usepackage{graphicx}
\usepackage{epstopdf}
\usepackage[cmex10]{amsmath}
\usepackage{subfigure}
\usepackage{amssymb}

\begin{document}

\title{Full-Duplex Relay based on Zero-Forcing Beamforming}
\author{Jong-Ho~Lee~and~Oh-Soon~Shin
\thanks{J.-H. Lee is with the Division of Electrical, Electronic \& Control Engineering, Kongju National University,
Korea (e-mail: jongholee@kongju.ac.kr).}
\thanks{O.-S. Shin is with the School of Electronic Engineering, Soongsil University, Seoul 156-743, Korea (e-mail: osshin@ssu.ac.kr).}
}

\maketitle

\begin{abstract}
In this paper, we propose a full-duplex relay (FDR) based on a zero-forcing beamforming (ZFBF) for a multiuser MIMO relay system. The ZFBF is employed at the base station to suppress both the self-interference of the relay and the multiuser interference at the same time. Numerical results show that the proposed FDR can enhance the sum rate performance as compared to the half-duplex relay (HDR), if sufficient isolation between the transmit and receive antennas is ensured at the relay.
\end{abstract}

\begin{IEEEkeywords}
Decode-and-forward, full-duplex relay (FDR), zero-forcing beamforming (ZFBF).
\end{IEEEkeywords}


\section{Introduction} \label{sec1}
\IEEEPARstart{R}{elay} transmission technology has attracted growing attention due to its promising capability of extending cell coverage and increasing cell capacity of wireless communication systems. So far, most relay technologies have been developed under the half-duplex constraint \cite{Laneman04} that prevents relays from transmitting and receiving at the same time. Although this form of relay, called half-duplex relay (HDR), is easy to implement, it requires partitioning of resource for transmission and reception, reducing the whole system capacity.

In order to overcome the drawback of HDR, we consider a full-duplex relay (FDR) that is able to transmit and receive on the same frequency at the same time \cite{Wang05}, where it is crucial to reduce the effect of possibly strong self-interference caused by its own transmitter to its own receiver. In order to solve the self-interference problem, \cite{Slingsby95} and \cite{Anderson04} have studied the physical isolation between the transmit and receive antennas at the RS. Since the antenna isolation may not be enough for stable FDR operation, \cite{Bliss07}-\cite{Ju09} have suggested signal processing techniques to suppress the residual self-interference. In particular, \cite{Bliss07} and \cite{Riihonen09} suggested spatial interference suppression schemes that exploit multiple transmit and receive antennas at the RS. However, the need for multiple antennas at the relay increases the complexity and cost of the relay, which may become a huge barrier for deploying relay stations in wireless networks. 

We propose an FDR that works under a multiuser MIMO scenario, where we need to suppress the multiuser interference as well as the self-interference in the FDR. We exploit a zero-forcing beamforming (ZFBF) in \cite{Yoo05} using multiple antennas at the base station (BS) rather than at the relay station (RS). It should be noted that the ZFBF is applied to cancel the self-interference of the FDR as well as to support multiuser MIMO. Moreover, the need for multiple antennas at the RS is eliminated by letting the BS perform the self-interference cancellation. The performance of the proposed FDR will be compared with that of an HDR with ZFBF and an ordinary multiuser MIMO without RS.

\section{System Model} \label{sec2}
We consider a downlink scenario where the BS has $L$ transmit antennas and a RS has one receive antenna and one transmit antenna. The RS is assumed to employ the decode-and-forward protocol \cite{Nabar04}, where the RS decodes all the signals from the BS and forwards them to the MS. We categorize mobile stations (MS's), each assumed to be equipped with one receive antenna, into two groups according to the reachability of the signals from the BS: 1) MS's in the first group can receive the signals from both the BS and RS. 2) MS's in the second group can receive only the signal from the RS. For notational convenience, we will refer to the set of $N_i$ MS's in the $i$-th group as MS-$i$, where $N_i$ denotes the number of MS's in the MS-$i$. Moreover, we assume that the BS selects the MS's from each MS group so that the sum rate is maximized \cite{Yoo05}.

The received signal $y_{i,j}$ at the $j$-th MS in the MS-$i$, $i = 1, 2$, is respectively given as
\begin{equation} \label{eq:received_signal}
\begin{array}{lcl}
{y}_{1,j} &=& \mathbf{h}_{1,j}^{\textrm{BS}}\mathbf{s}^{\textrm{BS}} + {h}_{1,j}^{\textrm{RS}}{s}^{\textrm{RS}} + {z}_{1,j}, \\
{y}_{2,j} &=& {h}_{2,j}^{\textrm{RS}}{s}^{\textrm{RS}} + {z}_{2,j},
\end{array}
\end{equation}
where $\mathbf{h}_{1,j}^{\textrm{BS}}$ denotes a $1 \times L$ channel vector between the BS and the $j$-th MS in MS-1. Moreover, ${h}_{i,j}^{\textrm{RS}}$ is a channel coefficient between the RS and the $j$-th MS in MS-$i$, $\mathbf{s}^{\textrm{BS}}$ denotes the $L \times 1$ transmit signal vector from the BS, ${s}^{\textrm{RS}}$ denotes the transmit signal from the RS, and ${z}_{i,j}$ denotes the zero-mean additive white Gaussian noise (AWGN) with unit variance. Note that $\mathbf{s}^{\textrm{BS}}$ and $s^{\textrm{RS}}$ are determined according to the MS selection. When generating $\mathbf{s}^{\textrm{BS}}$ at the BS and $s^{\textrm{RS}}$ at the RS, the total available transmit power is assumed to be ${P}_{T}^{\textrm{BS}}$ and ${P}_{T}^{\textrm{RS}}$, respectively. The signal transmission at the BS and RS is assumed to occur every time slot, the duration of which is assumed to be normalized to the unity.

\section{Relay Transmission Schemes} \label{sec3}
Although our main interest is in the FDR with ZFBF, the HDR transmission can also accommodate a ZFBF. For fair comparison between the HDR and FDR, we first develop a ZFBF for the HDR and the derived ZFBF will be incorporated into the FDR. In this section, we assume that all the exact channel state informations (CSI's) are available at the BS as in \cite{Marzetta06}. The impact of CSI feedback errors will be discussed in Section \ref{sec4}.

\subsection{Half-Duplex Relay (HDR)} \label{sec3a}
An HDR transmission scheme can be realized by time sharing of each slot between the BS and RS. Specifically, in the first $(1-t)$ fraction of each time slot, the BS can serve the RS as well as MS's in the MS-1 using a ZFBF scheme. The received signal at the RS during this period can be expressed as
\begin{equation} \label{eq:received_rs_hdr}
{y}_{\textrm{RS}} = \mathbf{h}_{\textrm{RS}}^{\textrm{BS}}\mathbf{s}^{\textrm{BS}} + {z}_{\textrm{RS}},
\end{equation}
where ${z}_{\textrm{RS}}$ also has unit variance. Let $\Gamma \subset \{ 1, 2, \cdots , N_{\textrm{1}} \}$ with $|\Gamma| < L$ be the set of MS's served by the BS, which is a subset of indices for the MS's in the MS-1. With $\Gamma=\{ \gamma_1, \gamma_2, \cdots, \gamma_{|\Gamma|} \}$, the received signals at the MS's associated with $\Gamma$, $\mathbf{y}_{\Gamma}=\left[ y_{1,\gamma_1}~y_{1,\gamma_2}~\cdots~y_{1,\gamma_{|\Gamma|}} \right]^{T}$, can be given as 
\begin{equation} \label{eq:received_ms1_hdr}
\mathbf{y}_{\Gamma} = \mathbf{H}\left(\Gamma \right)\mathbf{s}^{\textrm{BS}} + \mathbf{z}_{\Gamma},
\end{equation}
where the $|\Gamma| \times L $ channel matrix $\mathbf{H}\left(\Gamma \right)$ is given as
\begin{equation}
\mathbf{H}\left( \Gamma \right) = \left[ \left(\mathbf{h}^{\textrm{BS}}_{1,\gamma_1}\right)^{T}~ \left(\mathbf{h}^{\textrm{BS}}_{1,\gamma_2} \right)^{T}~ \cdots ~ \left(\mathbf{h}^{\textrm{BS}}_{1,\gamma_{|\Gamma|}}\right)^{T} \right]^{T}. \label{eq:channel_stack_Gamma}
\end{equation}
From (\ref{eq:received_rs_hdr}) and (\ref{eq:received_ms1_hdr}), we have
\begin{equation}\left[
  \begin{array}{c}
    \mathbf{y}_{\Gamma} \\
    {y}_{\textrm{RS}}
  \end{array}
\right] = \mathbf{H}_{\Gamma}
\mathbf{s}^{\textrm{BS}}+
\left[
  \begin{array}{c}
    \mathbf{z}_{\Gamma} \\
    {z}_{\textrm{RS}}
  \end{array}
\right], \label{eq:received_hdr_zf}
\end{equation}
where the $(|\Gamma|+1) \times L$ channel matrix $\mathbf{H}_{\Gamma}$ is defined as
\begin{equation}
\mathbf{H}_{\Gamma} \triangleq \left[
  \begin{array}{c}
    \mathbf{H}\left(\Gamma \right) \\
    \mathbf{h}_{\textrm{RS}}^{\textrm{BS}} \\
  \end{array}
\right], \label{eq:channel_hdr_zf}
\end{equation}
and the $L \times 1$ BS transmit signal vector $\mathbf{s}^{\textrm{BS}}_{\Gamma}$ can be expressed as
\begin{equation}
\mathbf{s}^{\textrm{BS}} \triangleq \mathbf{W}_{\Gamma} \mathbf{P}_{\Gamma} \left[
  \begin{array}{c}
    \mathbf{x}_{\Gamma} \\
    {x}_{\textrm{RS}} \\
  \end{array}
\right]. \label{eq:tx_signal_hdr_zf}
\end{equation}
In (\ref{eq:tx_signal_hdr_zf}), the $|\Gamma| \times 1$ transmit symbol vector $\mathbf{x}_{\Gamma}$ is defined as $\mathbf{x}_{\Gamma} = \left[ x_{1,\gamma_1}~x_{1,\gamma_2}~\cdots~x_{1,\gamma_{|\Gamma|}} \right]^{T}$, where $x_{1,\gamma_m}$ is the data symbol destined for the $\gamma_m$-th MS in the MS-1. ${x}_{\textrm{RS}}$ is the transmit symbol destined for the RS, which the RS decodes and forwards to an MS in the MS-2. The $L \times (|\Gamma|+1)$ ZFBF precoder $\mathbf{W}_{\Gamma}$ and the $(|\Gamma|+1) \times (|\Gamma|+1)$ power allocation matrix $\mathbf{P}_{\Gamma}$ are defined as
\begin{eqnarray}
\mathbf{W}_{\Gamma} &\triangleq& \mathbf{H}_{\Gamma}^{\dag}\left(\mathbf{H}_{\Gamma}\mathbf{H}_{\Gamma}^{\dag} \right)^{-1} \nonumber \\
&=&\left[
  \begin{array}{ccccc}
    \mathbf{w}^{\textrm{BS}}_{1,\gamma_1} & \mathbf{w}^{\textrm{BS}}_{1,\gamma_2} & \cdots & \mathbf{w}^{\textrm{BS}}_{1,\gamma_{|\Gamma|}} & \mathbf{w}^{\textrm{BS}}_{\textrm{RS}}
  \end{array}
\right] \label{eq:bf_vector}
\end{eqnarray}
and
\begin{equation}
{\mathbf{P}}_{\Gamma} = \mbox{diag} \left[\sqrt{{P}_{1,\gamma_1}}, \sqrt{{P}_{1,\gamma_2}}, \cdots, \sqrt{{P}_{1,\gamma_{|\Gamma|}}}, \sqrt{{P}_{\textrm{RS}}} \right] \label{eq:power_hdr1}
\end{equation}
with $(\cdot)^{\dag}$ denoting the conjugate transpose. The $L \times 1$ vectors $\mathbf{w}^{\textrm{BS}}_{1,\gamma_m}$ and $\mathbf{w}^{\textrm{BS}}_{\textrm{RS}}$ in (\ref{eq:bf_vector}) denote the ZFBF vectors at the BS corresponding to the $\gamma_m$-th MS in the MS-1 and to the RS, respectively. Then, the sum rate for each link during the $(1-t)$ fraction of a time slot is given as
\begin{equation}
\begin{array}{lcl}
{R}_{\Gamma} &=& \sum_{m=1}^{|\Gamma|} {\log}_{2} \left( 1+{P}_{1,\gamma_m} \right), \\
{R}_{\textrm{RS}} &=& {\log}_{2} \left( 1+{P}_{\textrm{RS}} \right) 
\end{array}\label{eq:cap_hdr_zf1}
\end{equation}
where the transmit power ${P}_{1,\gamma_m}$ allocated to the $\gamma_m$-th MS in the MS-1 and $P_{\textrm{RS}}$ allocated to the RS should satisfy the following power constraint,
\begin{equation}
\sum_{m =1}^{|\Gamma|} ||\mathbf{w}^{\textrm{BS}}_{1,\gamma_m}||^2 {P}_{1,\gamma_m} + ||\mathbf{w}^{\textrm{BS}}_{\textrm{RS}}||^2 {P}_{\textrm{RS},\Gamma} = {P}_{T}^{\textrm{BS}}. \label{eq:bs_pw_hdr_zf}
\end{equation}
Here, we define $\tau_{1,\gamma_m}=\frac{1}{||\mathbf{w}^{\textrm{BS}}_{1,\gamma_m}||^2}$ and $\tau_{\textrm{RS}}=\frac{1}{||\mathbf{w}^{\textrm{BS}}_{\textrm{RS}}||^2}$, and perform a water-filling as in \cite{Yoo05}. Then, ${P}_{1,\gamma_m}$ and $P_{\textrm{RS}}$ that maximize the sum rate ${R}_{\Gamma}+{R}_{\textrm{RS}}$ are given as
\begin{equation}
\begin{array}{lcl}
{P}_{1,\gamma_m} &=& \left(\mu \tau_{1,\gamma_m} -1 \right)^{+}, \\
{P}_{\textrm{RS}} &=& \left(\mu \tau_{\textrm{RS}} -1 \right)^{+} 
\end{array}\label{eq:water_filling_hdr_zf1}
\end{equation}
where $(x)^{+}$ denotes $\max\{x,0 \}$ and 
\begin{equation}
\sum_{m =1}^{|\Gamma|} \left(\mu - \frac{1}{\tau_{1,\gamma_m}} \right)^{+} + \left(\mu - \frac{1}{ \tau_{\textrm{RS}}} \right)^{+} = {P}_{T}^{\textrm{BS}}.
\end{equation}

In the remaining $t$ fraction of the time slot, we consider two possible scenarios: 1) the BS serves MS's in the MS-1 and the RS forwards the decoded data to an MS in the MS-2 at the same time. 2) the BS serves MS's in the MS-1 and the RS forwards the decoded data to another MS in the MS-1 at the same time. For the first scenario, the RS's signal destined for an MS in the MS-2 becomes the multiuser interference to the MS's in the MS-1. In the following, we cancel this interference by using a ZFBF, which will be incorporated into the FDR in the next subsection. The received signals at the MS's in the MS-1 and an MS in the MS-2 can be expressed as
\begin{equation}\left[
  \begin{array}{c}
    {\mathbf{y}}_{\check{\Gamma}} \\
    {y}_{2,j} \\
  \end{array}
\right] = \mathbf{H}_{\check{\Gamma},j}\left[
  \begin{array}{c}
    \mathbf{s}^{\textrm{BS}}\\
    {s}^{\textrm{RS}}
  \end{array}
\right]
+
\left[
  \begin{array}{c}
    {\mathbf{z}}_{\check{\Gamma}} \\
    {z}_{2,j} \\
  \end{array}
\right]. \label{eq:received_hdr_zf2}
\end{equation}
The $(|\check{\Gamma}|+1) \times (L+1)$ channel matrix ${\mathbf{H}}_{\check{\Gamma}}$ is given as
\begin{equation}
{\mathbf{H}}_{\check{\Gamma},j}=\left[
  \begin{array}{cc}
    \mathbf{H}\left( \check{\Gamma} \right) & \mathbf{h}\left(\check{\Gamma} \right) \\
    \mathbf{0}_{1 \times L} & {h}_{2,j}^{\textrm{RS}} \\
  \end{array}
\right], \label{eq:channel_hdr_zf2}
\end{equation}
where $\mathbf{H}\left( \check{\Gamma} \right)$ can be defined as in (\ref{eq:channel_stack_Gamma}) and $\mathbf{h}\left(\check{\Gamma} \right)=\left[{h}_{1,\check{\gamma}_1}^{\textrm{RS}}~{h}_{1,\check{\gamma}_2}^{\textrm{RS}}~\cdots~{h}_{1,\check{\gamma}_{|\check{\Gamma}|}}^{\textrm{RS}} \right]^{T}$. The $(L+1) \times 1$ transmit signal vector can be expressed in a matrix form
\begin{equation}
\left[
  \begin{array}{c}
    \mathbf{s}^{\textrm{BS}}\\
    {s}^{\textrm{RS}}
  \end{array}
\right] = {\mathbf{W}}_{\check{\Gamma},j}
{\mathbf{P}}_{\check{\Gamma},j}
\left[
  \begin{array}{c}
    \mathbf{x}_{\check{\Gamma}} \\
    \tilde{x}_{2,j} \\
  \end{array}
\right] \label{eq:transmit_hdr_zf2}
\end{equation}
with $\mathbf{x}_{\check{\Gamma}} = \left[ x_{1,\check{\gamma}_1}~x_{1,\check{\gamma}_2}~\cdots~x_{1,\gamma_{|\check{\gamma}|}} \right]^{T}$, $\check{\Gamma}=\{ \check{\gamma}_1~\check{\gamma}_2~\cdots~\check{\gamma}_{|\check{\Gamma}|} \}$ and $|\check{\Gamma}| \leq L$. In (\ref{eq:transmit_hdr_zf2}), the $(L+1) \times (|\check{\Gamma}|+1)$ ZFBF precoder and the $(|\check{\Gamma}|+1) \times (|\check{\Gamma}|+1)$ power allocation matrix are given as
\begin{equation}
{\mathbf{W}}_{\check{\Gamma},j} = {\mathbf{H}}_{\check{\Gamma},j}^{\dag}\left({\mathbf{H}}_{\check{\Gamma},j}{\mathbf{H}}_{\check{\Gamma},j}^{\dag} \right)^{-1} \label{eq:precode_hdr_zf2}
\end{equation}
and
\begin{equation}
{\mathbf{P}}_{\check{\Gamma},j} = \mbox{diag} \left[\sqrt{{P}_{1,\check{\gamma}_1}}, \sqrt{{P}_{1,\check{\gamma}_2}}, \cdots, \sqrt{{P}_{1,\check{\gamma}_{|\check{\Gamma}|}}},\sqrt{{P}_{2,j}} \right], \label{eq:power_hdr2}
\end{equation}
respectively. Note that $\tilde{x}_{2,j}$ denotes a delayed version of ${x}_{2,j}$ destined for the $j$-th MS in the MS-2 with a delay corresponding to the decoding delay at the RS. ${\mathbf{W}}_{\check{\Gamma}}$ in (\ref{eq:precode_hdr_zf2}) can be rewritten as \cite{Frank90}
\begin{equation}
{\mathbf{W}}_{\check{\Gamma},j} = \left[
  \begin{array}{cc}
    {\mathbf{W}}_{\check{\Gamma}}^{\textrm{BS}} & {\mathbf{w}}_{2,j}^{\textrm{BS}}  \\
    \mathbf{0}_{1 \times |\check{\Gamma}|} & {{w}}_{2,j}^{\textrm{RS}}
  \end{array}
\right] \label{eq:precode1_hdr_zf2}
\end{equation}
where ${{w}}_{2,j}^{\textrm{RS}}$ is the beamforming weight at the RS and ${\mathbf{w}}_{2,j}^{\textrm{BS}}$ denotes the $L \times 1$ ZFBF vectors at the BS corresponding to the $j$-th MS in the MS-2. The $L \times |\check{\Gamma}|$ beamforming matrix ${\mathbf{W}}_{\check{\Gamma}}^{\textrm{BS}}$ is defined as
\begin{equation}
{\mathbf{W}}_{\check{\Gamma}}^{\textrm{BS}} = \left[
  \begin{array}{cccc}
    {\mathbf{w}}_{1,\check{\gamma}_1}^{\textrm{BS}} & {\mathbf{w}}_{1,\check{\gamma}_2}^{\textrm{BS}} & \cdots & {\mathbf{w}}_{1,\check{\gamma}_{|\check{\Gamma}|}}^{\textrm{BS}}
  \end{array}
\right].
\end{equation}
Then, the transmit signals in (\ref{eq:transmit_hdr_zf2}) can be rewritten as
\begin{eqnarray}
\mathbf{s}^{\textrm{BS}} &=& \sum_{m=1}^{|\check{\Gamma}|} {\mathbf{w}}_{1,{\check{\gamma}}_m}^{\textrm{BS}}\sqrt{{P}_{1,\check{\gamma}_m}}{x}_{1,\check{\gamma}_m}+\mathbf{w}_{2,j}^{\textrm{BS}}\sqrt{{P}_{2,j}} \tilde{x}_{2,j}, \nonumber \\
{s}^{\textrm{RS}} &=& {w}_{2,j}^{\textrm{RS}} \sqrt{{P}_{2,j}} \tilde{x}_{2,j}.
\label{eq:transmit_hdr_zf2_2}
\end{eqnarray}
Note that the BS transmit signal $\mathbf{s}^{\textrm{BS}}$ contains not only the beamformed signal of $\mathbf{x}_{\check{\Gamma}}$ but also the beamformed signal of $\tilde{x}_{2,j}$. It should be noted that the BS can transmit the beamformed signal of $\tilde{x}_{2,j}$, since $\tilde{x}_{2,j}$ is a delayed version of ${x}_{2,j}$ that is originated from the BS. It is also seen that the RS transmits only the beamformed signal of $\tilde{x}_{2,j}$, due to the form of the precoding matrix ${\mathbf{W}}_{\check{\Gamma},j}$ in (\ref{eq:precode1_hdr_zf2}). It is reasonable because the RS can know only $\tilde{x}_{2,j}$ after decoding ${x}_{2,j}$.

The sum rate served to MS's is given as
\begin{equation}
\begin{array}{lcl}
{R}_{\check{\Gamma}} &=& \sum_{m=1}^{|\check{\Gamma}|}\log_{2} \left( 1+{P}_{1,\check{\gamma}_m} \right), \\
{R}_{2,j} &=& \log_{2} \left( 1+{P}_{2,j} \right). \label{eq:cap_hdr_zf2}
\end{array}
\end{equation}
From (\ref{eq:transmit_hdr_zf2_2}), it is obvious that the BS power constraint is given as
\begin{equation}
\sum_{m=1}^{|\check{\Gamma}|}|| {\mathbf{w}}_{1,\check{\gamma}_m}^{\textrm{BS}} ||^2 {P}_{1,\check{\gamma}_m} + || {\mathbf{w}}_{2,j}^{\textrm{BS}} ||^2  {P}_{2,j} = {P}_{T}^{\textrm{BS}}. \label{eq:bs_pw_hdr}
\end{equation}
As in (\ref{eq:water_filling_hdr_zf1}), the water-filling provides the available power $\bar{P}_{1,\check{\gamma}_m}^{\textrm{BS}}$ and $\bar{P}_{2,j}^{\textrm{BS}}$ at the BS. From the RS power constraint, the available power at the RS can be also computed as
\begin{equation}
\bar{P}_{2,j}^{\textrm{RS}}  = \frac{{P}_{T}^{\textrm{RS}}}{\left| {{w}}_{2,j}^{\textrm{RS}} \right|^{2}}.
\end{equation}
Now, we determine final transmit power ${P}_{1,\check{\gamma}_m}^{\star}$ and ${P}_{2,j}^{\star}$ using $\bar{P}_{1,\check{\gamma}_m}^{\textrm{BS}}$, $\bar{P}_{2,j}^{\textrm{BS}}$ and $\bar{P}_{2,j}^{\textrm{RS}}$. In the case of $\bar{P}_{2,j}^{\textrm{RS}} > \bar{P}_{2,j}^{\textrm{BS}}$, we simply set ${P}_{1,\check{\gamma}_m}^{\star} = \bar{P}_{1,\check{\gamma}_m}^{\textrm{BS}}$ and ${P}_{2,j}^{\star} = \bar{P}_{2,j}^{\textrm{BS}}$. Otherwise, we set ${P}_{2,j}^{\star} = \bar{P}_{2,j}^{RS}$. Then, we have
\begin{equation}
\sum_{m=1}^{|\check{\Gamma}|}|| {\mathbf{w}}_{1,\check{\gamma}_m}^{\textrm{BS}} ||^2 {P}_{1,\check{\gamma}_k}  = {P}_{T}^{\textrm{BS}}- || {\mathbf{w}}_{2,j}^{\textrm{BS}} ||^2  {P}_{2,j}^{\star}
\end{equation}
and perform the water-filling again to determine ${P}_{1,\check{\gamma}_m}^{\star}$. Then, ${P}_{1,\check{\gamma}_m}^{\star}$ and ${P}_{2,j}^{\star}$ are used to compute (\ref{eq:cap_hdr_zf2}). Assuming that all the received signals in the $(1-t)$ time fraction are decoded and forwarded by the RS during the remaining $t$ time fraction, $t$ is chosen such that $(1-t){R}_{\textrm{RS}} = t{R}_{2,j}$ \cite{Chae08}. This implies that the RS can forward only the information that has been received from the BS. Then, the overall sum rate in the HDR is given as
\begin{eqnarray}
{R}_{\textrm{HDR}} = \max_{\begin{smallmatrix} {\Gamma}, \check{\Gamma} \subset \{ 1, 2, \cdots , N_{1} \} \\ j \in \{ 1, 2, \cdots , N_2 \} \end{smallmatrix}} \frac{{R}_{\Gamma} {R}_{2,j} + R_{\textrm{RS}}({R}_{\check{\Gamma}}+{R}_{2,j})}{{R}_{2,j}+R_{\textrm{RS}}}. \label{eq:cap_hdr_zf}
\end{eqnarray}

For the second scenario, the RS receives ${x}_{1,j}$ from the BS and forwards $\tilde{x}_{1,j}$ to the $j$-th MS in the MS-1. ${\mathbf{H}}_{\check{\Gamma},j}$ can be given as
\begin{equation}
{\mathbf{H}}_{\check{\Gamma},j}=\left[
  \begin{array}{cc}
    \mathbf{H}\left( \check{\Gamma} \right) & \mathbf{h}\left(\check{\Gamma} \right) \\
    \mathbf{h}^{\textrm{BS}}_{1,j} & {h}_{1,j}^{\textrm{RS}} \\
  \end{array}
\right]. \label{eq:channel_hdr_zf2_na}
\end{equation}
Here, the MS-1 receives the signals from both the BS and RS such that $\mathbf{h}^{\textrm{BS}}_{1,j}$ is not a zero vector. The pseudo-inverse of (\ref{eq:channel_hdr_zf2_na}) yields
\begin{equation}
{\mathbf{W}}_{\check{\Gamma},j} = \left[
  \begin{array}{cc}
    {\mathbf{W}}_{\check{\Gamma}}^{\textrm{BS}} & {\mathbf{w}}_{1,j}^{\textrm{BS}}  \\
    {\mathbf{w}}_{1,j}^{\textrm{RS}} & {{w}}_{1,j}^{\textrm{RS}}
  \end{array}
\right], \label{eq:precode1_hdr_zf2_na}
\end{equation}
where the $1 \times |\check{\Gamma}|$ vector ${\mathbf{w}}_{1,j}^{\textrm{RS}}$ is also not a zero vector unlike in (\ref{eq:precode1_hdr_zf2}). From (\ref{eq:transmit_hdr_zf2}), we obtain the RS transmit signal shown as
\begin{eqnarray}
{s}^{\textrm{RS}} &=& {\mathbf{w}}_{1,j}^{\textrm{RS}}\mbox{diag} \left[\sqrt{{P}_{1,\check{\gamma}_1}}, \sqrt{{P}_{1,\check{\gamma}_2}}, \cdots, \sqrt{{P}_{1,\check{\gamma}_{|\check{\Gamma}|}}} \right]\mathbf{x}_{\check{\Gamma}} \nonumber\\ &&+{{w}}_{1,j}^{\textrm{RS}}\sqrt{{P}_{1,j}}\tilde{x}_{1,j}. \label{eq:transmit_hdr_zf2_na}
\end{eqnarray}
Note that the RS has no information on $\mathbf{x}_{\check{\Gamma}}$ since it received and decoded only ${x}_{1,j}$ in the $(1-t)$ time fraction. Therefore, it is obvious that the RS cannot compose (\ref{eq:transmit_hdr_zf2_na}), and thus we do not consider this transmission scenario. Due to the same reason, the scenario cannot be realized in the FDR as well, which will be described in the next subsection.

\subsection{Full-Duplex Relay (FDR)} \label{sec3c}
In the FDR transmission, the received signal at the RS and at an MS in each group can be expressed as
\begin{equation}\label{eq:received_rs_fdr}
\begin{array}{lcl}
{y}_{1,j} &=& \mathbf{h}_{1,j}^{\textrm{BS}}\mathbf{s}^{\textrm{BS}} + {h}_{1,j}^{\textrm{RS}}{s}^{\textrm{RS}} + {z}_{1,j}, \\
{y}_{2,j} &=& {h}_{2,j}^{\textrm{RS}}{s}^{\textrm{RS}} + {z}_{2,j}, \\
{y}_{\textrm{RS}} &=& \mathbf{h}_{\textrm{RS}}^{\textrm{BS}}\mathbf{s}^{\textrm{BS}}+{h}_{\textrm{RS}}^{\textrm{RS}}{s}^{\textrm{RS}}+{z}_{\textrm{RS}},
\end{array}
\end{equation}
where the second term in ${y}_{\textrm{RS}}$ represents the self-interference due to simultaneous transmission and reception. During an entire time slot, the BS serves MS's in the MS-1 as well as the RS, while the RS serves an MS in the MS-2 at the same time. In the proposed FDR, ${h}_{\textrm{RS}}^{\textrm{RS}}$ should be available at the BS. Here, we assume that the RS transmits ${h}_{\textrm{RS}}^{\textrm{RS}}$ to the BS by using an appropriate method, such as the analog linear modulation in \cite{Marzetta06}.

With $\Gamma \subset \{ 1, 2, \cdots , N_{1} \}$ and $|\Gamma| < L$, the received signals can be arranged in a matrix form as
\begin{equation}
\left[
  \begin{array}{c}
    \mathbf{y}_{\Gamma} \\
    {y}_{\textrm{RS}} \\
    {y}_{2,j} \\
  \end{array}
\right] = {\mathbf{H}}_{\Gamma,j}
\left[
  \begin{array}{c}
    \mathbf{s}^{\textrm{BS}}\\
    {s}^{\textrm{RS}}
  \end{array}
\right]
+
\left[
  \begin{array}{c}
    \mathbf{z}_{\Gamma} \\
    {z}_{\textrm{RS}} \\
    {z}_{2,j} \\
  \end{array}
\right], \label{eq:received_fdr_zf}
\end{equation}
where the $(|\Gamma|+2) \times (L+1)$ channel matrix ${\mathbf{H}}_{\Gamma,j}$ is given as
\begin{equation}
{\mathbf{H}}_{\Gamma,j} = \left[
  \begin{array}{cc}
    \mathbf{H}\left(\Gamma \right) & \mathbf{h}\left(\Gamma \right)  \\
    \mathbf{h}_{\textrm{RS}}^{\textrm{BS}} & {h}_{\textrm{RS}}^{\textrm{RS}} \\
    \mathbf{0}_{1 \times L} & {h}_{2,j}^{\textrm{RS}} \\
  \end{array}
\right] \label{eq:channel_fdr_zf}
\end{equation}
where $\mathbf{H}\left(\Gamma \right)$ is defined in (\ref{eq:channel_stack_Gamma}) and $\mathbf{h}\left(\Gamma \right)=\left[{h}_{1,\gamma_1}^{\textrm{RS}}~{h}_{1,\gamma_2}^{\textrm{RS}}~\cdots~{h}_{1,\gamma_{|\Gamma|}}^{\textrm{RS}} \right]^{T}$. The $L \times 1$ BS transmit signal vector $\mathbf{s}^{\textrm{BS}}$ and the RS transmit signal ${s}^{\textrm{RS}}$ are expressed in a matrix form as
\begin{equation}
\left[
  \begin{array}{c}
    \mathbf{s}^{\textrm{BS}}\\
    {s}^{\textrm{RS}}
  \end{array}
\right]=\mathbf{W}_{\Gamma,j}
\mathbf{P}_{\Gamma,j}
\left[
  \begin{array}{c}
    \mathbf{x}_{\Gamma} \\
    {x}_{2,j} \\
    \tilde{x}_{2,j} \\
  \end{array}
\right]. \label{eq:tx_signal_fdr_zf}
\end{equation}
In (\ref{eq:tx_signal_fdr_zf}), $\mathbf{x}_{\Gamma}$ and ${x}_{2,j}$ denote transmit symbols of the BS destined for the MS's in the MS-1 and for the $j$-th MS in the MS-2, respectively, and $\tilde{x}_{2,j}$ denotes transmit symbol of the RS. Moreover, the $(L+1) \times (|\Gamma|+2)$ ZFBF precoder and the $(|\Gamma|+2) \times (|\Gamma|+2)$ power allocation matrix in (\ref{eq:tx_signal_fdr_zf}) are defined as
\begin{equation}
\mathbf{W}_{\Gamma,j} = \mathbf{H}_{\Gamma,j}^{\dag}\left(\mathbf{H}_{\Gamma,j}\mathbf{H}_{\Gamma,j}^{\dag} \right)^{-1} \label{eq:precode_fdr_zf}
\end{equation}
and
\begin{equation}
\mathbf{P}_{\Gamma,j} = \mbox{diag} \left[\sqrt{{P}_{1,\gamma_1}}, \sqrt{{P}_{1,\gamma_2}}, \cdots , \sqrt{{P}_{1,\gamma_{|\Gamma|}}}, \sqrt{{P}_{\textrm{RS}}}, \sqrt{{P}_{2,j}} \right], \label{eq:power_fdr}
\end{equation}
respectively. In (\ref{eq:power_fdr}), $P_{1,\gamma_m}$ is the BS power allocated to the $\gamma_m$-th MS in the MS-1, $P_{\textrm{RS}}$ is the BS power allocated to the RS, and ${P}_{2,j}$ is the RS power allocated to the $j$-th MS in the MS-2. 

The ZFBF precoder in (\ref{eq:precode_fdr_zf}) can be rewritten as \cite{Frank90}
\begin{equation}
\mathbf{W}_{\Gamma,j} = \left[
  \begin{array}{ccc}
    \mathbf{W}_{\Gamma}^{\textrm{BS}} & \mathbf{w}_{\textrm{RS}}^{\textrm{BS}} & \mathbf{w}_{2,j}^{\textrm{BS}} \\
    \mathbf{0}_{1 \times |\Gamma|} & 0 & {w}_{2,j}^{\textrm{RS}} \\
  \end{array}
\right], \label{eq:precode1_fdr_zf}
\end{equation}
where $\mathbf{w}_{\textrm{RS}}^{\textrm{BS}}$ and $\mathbf{w}_{2,j}^{\textrm{BS}}$ denote the $L \times 1$ ZFBF vectors at the BS, ${w}_{2,j}^{\textrm{RS}}$ is the beamforming weight at the RS and
\begin{equation}
\mathbf{W}_{\Gamma}^{\textrm{BS}} = \left[
  \begin{array}{cccc}
    {\mathbf{w}}_{1,{\gamma}_1}^{\textrm{BS}} & {\mathbf{w}}_{1,{\gamma}_2}^{\textrm{BS}} & \cdots & {\mathbf{w}}_{1,{\gamma}_{|\Gamma|}}^{\textrm{BS}}
  \end{array}
\right].
\end{equation}
Then, the transmit signals in (\ref{eq:tx_signal_fdr_zf}) can be rewritten as
\begin{eqnarray}
\mathbf{s}^{\textrm{BS}} &=& \sum_{m=1}^{|\Gamma|} {\mathbf{w}}_{1,{\gamma}_m}^{\textrm{BS}}\sqrt{{P}_{1,\gamma_m}}{x}_{1,\gamma_m}+\mathbf{w}_{\textrm{RS}}^{\textrm{BS}}\sqrt{{P}_{\textrm{RS}}} {x}_{2,j}\nonumber \\ &&+\mathbf{w}_{2,j}^{\textrm{BS}}\sqrt{{P}_{2,j}} \tilde{x}_{2,j}, \nonumber\\
{s}^{\textrm{RS}} &=& {w}_{2,j}^{\textrm{RS}} \sqrt{{P}_{2,j}} \tilde{x}_{2,j}.
\end{eqnarray}
As described in Section \ref{sec3a}, it is reasonable that $\mathbf{s}^{\textrm{BS}}$ contains not only the beamformed signal of $\mathbf{x}_{\Gamma}$ and $x_{2,j}$ but also the beamformed signal of $\tilde{x}_{2,j}$. Moreover, the RS transmits only the beamformed signal of $\tilde{x}_{2,j}$.

Substituting (\ref{eq:tx_signal_fdr_zf})-(\ref{eq:power_fdr}) into (\ref{eq:received_fdr_zf}), we get
\begin{equation}
{y}_{1,\gamma_m}=\sqrt{{P}_{1,\gamma_m}} {x}_{1,\gamma_m}+{z}_{1,\gamma_m} \label{eq:afterzfbf_fdr_zf1}
\end{equation}
with $m=1,2,\cdots,|\Gamma|$ and
\begin{equation}
\left[
  \begin{array}{c}
    {y}_{\textrm{RS}} \\
    {y}_{2,j} \\
  \end{array}
\right] =
\left[
  \begin{array}{c}
    \sqrt{{P}_{\textrm{RS}}} {x}_{2,j} \\
    \sqrt{{P}_{2,j}} \tilde{x}_{2,j} \\
  \end{array}
\right]
+
\left[
  \begin{array}{c}
    {z}_{\textrm{RS}} \\
    {z}_{2,j} \\
  \end{array}
\right]. \label{eq:afterzfbf_fdr_zf2}
\end{equation}
In (\ref{eq:afterzfbf_fdr_zf2}), it is seen that the ZFBF precoder $\mathbf{W}_{\Gamma,j}$ makes ${y}_{\textrm{RS}}$ free from the self-interference. Thus, the RS can decode ${x}_{2,j}$ and forward $\tilde{x}_{2,j}$ to the $j$-th MS in the MS-2 without interference. It is also obvious that we can set ${P}_{\textrm{RS}} = {P}_{2,j}$, since ${x}_{2,j}$ and $\tilde{x}_{2,j}$ are associated with the same information rate. 

The sum rate served to MS's is found as
\begin{equation}
{R}_{\textrm{FDR}} = \max_{\begin{smallmatrix} \Gamma \subset \{1, 2, \cdots , N_{1} \} \\ j \in {1, 2, \cdots, N_2} \end{smallmatrix}} \sum_{m=1}^{|\Gamma|}\log_{2} \left( 1+{P}_{1,\gamma_m} \right) + \log_{2} \left( 1+{P}_{\textrm{RS}} \right) \label{eq:cap_fdr_zf}
\end{equation}
with the BS power constraint
\begin{equation}
\sum_{m=1}^{|\Gamma|} || \mathbf{w}_{1,{\gamma}_m}^{\textrm{BS}} ||^2 {P}_{1,{\gamma}_m} + \left( || \mathbf{w}_{\textrm{RS}}^{\textrm{BS}} ||^2 + || \mathbf{w}_{2,j}^{\textrm{BS}} ||^2 \right) {P}_{\textrm{RS}} = {P}_{T}^{\textrm{BS}}. \label{eq:bs_pw_fdr}
\end{equation}
After water-filling with (\ref{eq:cap_fdr_zf}) and (\ref{eq:bs_pw_fdr}) as in (\ref{eq:water_filling_hdr_zf1}), the available power $\bar{P}_{1,{\gamma}_m}$ and $\bar{P}_{\textrm{RS}}$ at the BS can be obtained. Similarly, the available power $\bar{P}_{2,j}$ at the RS can be computed as
\begin{equation}
\bar{P}_{2,j}  = \frac{{P}_{T}^{\textrm{RS}}}{\left|w_{2,j}^{\textrm{RS}} \right|^2}. \label{eq:rs_pw_fdr}
\end{equation}
Using $\bar{P}_{1,{\gamma}_m}$, $\bar{P}_{\textrm{RS}}$, and $\bar{P}_{2,j}$ obtained from (\ref{eq:bs_pw_fdr}) and (\ref{eq:rs_pw_fdr}), we determine final transmit power ${P}_{1,{\gamma}_m}^{\star}$ and ${P}_{\textrm{RS}}^{\star}$ of the BS. In the case of $\bar{P}_{\textrm{RS}} < \bar{P}_{2,j}$, we simply get ${P}_{1,{\gamma}_m}^{\star} = \bar{P}_{1,{\gamma}_m}$ and ${P}_{\textrm{RS}}^{\star} = \bar{P}_{\textrm{RS}}$. Otherwise, we set ${P}_{\textrm{RS}}^{\star} = \bar{P}_{2,j}$. Then we have
\begin{equation}
\sum_{m=1}^{|\Gamma|} || \mathbf{w}_{1,{\gamma}_m}^{\textrm{BS}} ||^2 {P}_{1,{\gamma}_m}  = {P}_{T}^{\textrm{BS}}- \left( || \mathbf{w}_{\textrm{RS}}^{\textrm{BS}} ||^2 + || \mathbf{w}_{2,j}^{\textrm{BS}} ||^2 \right) {P}_{\textrm{RS}}^{\star} \label{eq:bs_pw_fdr2}
\end{equation}
and ${P}_{1,{\gamma}_m}^{\star}$ is computed by the water-filling with (\ref{eq:bs_pw_fdr2}). Then, ${P}_{1,{\gamma}_m}^{\star}$ and ${P}_{\textrm{RS}}^{\star}$ are used to compute the sum rate of the proposed FDR in (\ref{eq:cap_fdr_zf}). In order to realize the FDR with ZFBF, the CSI between the RS's transmit antenna and receive antenna, $h_{\textrm{RS}}^{\textrm{RS}}$, is required at the BS. Moreover, the BS should inform the RS of $w_{2,j}^{\textrm{RS}} \sqrt{{P}_{\textrm{RS}}^{\star}}$ after constructing $\mathbf{W}_{\Gamma,j}$, ${P}_{1,{\gamma}_m}^{\star}$ and ${P}_{\textrm{RS}}^{\star}$.

\begin{figure*}[t]
 \centering \subfigure[]
  {\includegraphics[height=3in]{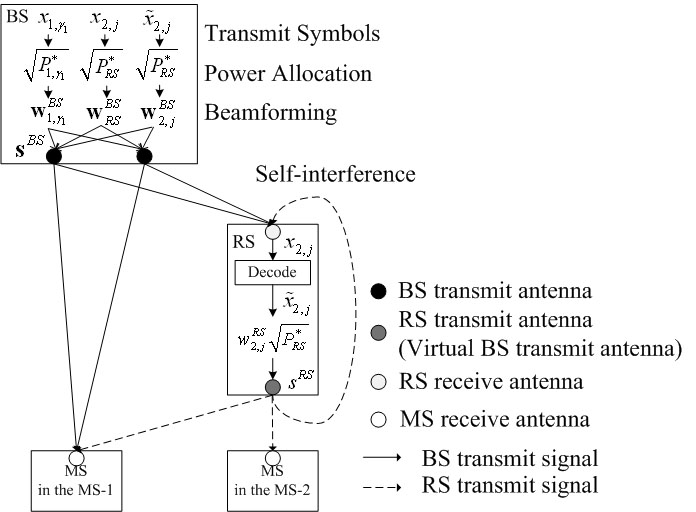}} \label{fig1a}\\
 \centering\subfigure[]
  {\includegraphics[height=3in]{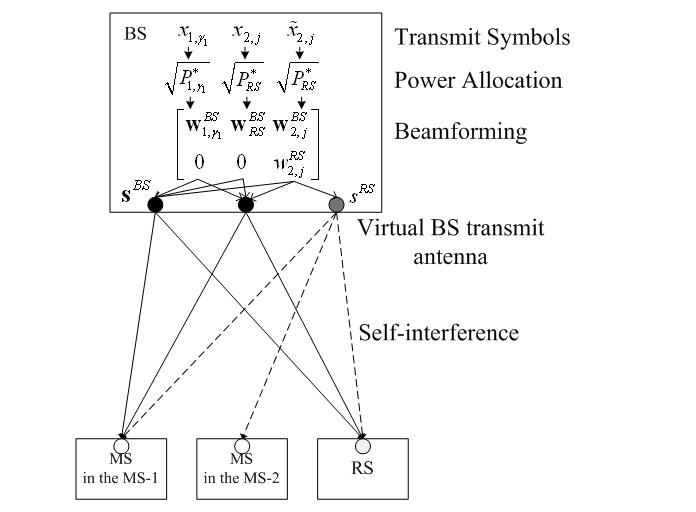}} \label{fig1b}
 \caption{FDR signal model and its equivalent model ($L=2$ and $|\Gamma|=1$). (a) Signal model for the proposed FDR. (b) The equivalent multiuser MIMO transmission model.}\label{fig1}
\end{figure*}

We can establish an interesting connection between the proposed FDR and multiuser MIMO transmission. The signal model for the proposed FDR is illustrated in Fig. 1(a), where the number of BS transmit antennas is assumed to be $L=2$. If we regard the RS transmit antenna as a virtual BS transmit antenna in Fig. 1(a), the FDR signal model can equivalently be translated into a multiuser MIMO transmission model in Fig. 1(b), where the restriction on the received signals at the MS-2 is due to our assumption of the signal reachability in Section \ref{sec2}. Comparing Figs. 1(a) and 1(b), we see that the self-interference in the FDR corresponds to multiuser interference in the equivalent multiuser MIMO model. This justifies our approach of eliminating both the self-interference and the multiuser interference simultaneously, since the ZFBF was adopted to suppress the multiuser interference in the multiuser MIMO system \cite{Yoo05}.

It is worth discussing a practical issue to be considered in deploying the FDR. The self-interference signal might be much stronger than the desired signal at the receiver front-end of the RS. In order to avoid possible saturation in this case, it is important to make sure that the received signals at the RS fall within the dynamic range of the receiver. This can be assured through the use of an analog-to-digital converter (ADC) with high resolution and/or sufficient isolation between the transmit and receive antennas \cite{Slingsby95}, \cite{Anderson04}. 

\section{Impact of CSI Feedback Errors} \label{sec4}
In order to investigate the impact of CSI feedback errors on the sum rate performance, we adopt the CSI error model in \cite{Yoo06}. According to the model, the CSI's at the BS can be represented as
\begin{equation}
\begin{array}{lcl}
\hat{\mathbf{h}}_{1,j}^{\textrm{BS}} &=& \mathbf{h}_{1,j}^{\textrm{BS}} + {e}_{1,j}^{\textrm{BS}}, \\
\hat{h}_{1,j}^{\textrm{RS}} &=& {h}_{1,j}^{\textrm{RS}}+{e}_{1,j}^{\textrm{RS}},\\
\hat{\mathbf{h}}_{\textrm{RS}}^{\textrm{BS}} &=& \mathbf{h}_{\textrm{RS}}^{\textrm{BS}}+{e}_{\textrm{RS}}^{\textrm{BS}},\\
\hat{h}_{\textrm{RS}}^{\textrm{RS}} &=& {h}_{\textrm{RS}}^{\textrm{RS}}+{e}_{\textrm{RS}}^{\textrm{RS}}, \\
\hat{h}_{2,j}^{\textrm{RS}} &=& {h}_{2,j}^{\textrm{RS}} + {e}_{2,j}^{\textrm{RS}},
\end{array}\label{eq:channel_fdr_zf_err}
\end{equation}
where ${e}_{1,j}^{\textrm{BS}}$, ${e}_{1,j}^{\textrm{RS}}$, ${e}_{2,j}^{\textrm{RS}}$, ${e}_{\textrm{RS}}^{\textrm{BS}}$, and ${e}_{\textrm{RS}}^{\textrm{RS}}$ denote the feedback errors, which are assumed to follow independent and identically distributed (i.i.d.) complex Gaussian distribution with zero mean and variance of $\sigma_{E}^{2}$. 

As in (\ref{eq:channel_fdr_zf}), we can compose $\hat{\mathbf{H}}_{\Gamma,j}$ using (\ref{eq:channel_fdr_zf_err}) and the FDR ZFBF precoder with the feedback errors, $\hat{\mathbf{W}}_{\Gamma,j} = \hat{\mathbf{H}}_{\Gamma,j}^{\dag}\left(\hat{\mathbf{H}}_{\Gamma,j}\hat{\mathbf{H}}_{\Gamma,j}^{\dag} \right)^{-1}$, can be written as 
\begin{equation}
\hat{\mathbf{W}}_{\Gamma,j} = \left[
  \begin{array}{ccccc}
    \hat{\mathbf{w}}_{1,{\gamma}_1}^{\textrm{BS}} & \cdots & \hat{\mathbf{w}}_{1,{\gamma}_{|\Gamma|}}^{\textrm{BS}} & \hat{\mathbf{w}}_{\textrm{RS}}^{\textrm{BS}} & \hat{\mathbf{w}}_{2,j}^{\textrm{BS}} \\
    0 & \cdots & 0 & 0 & \hat{w}_{2,j}^{\textrm{RS}} \\
  \end{array}
\right]. \label{eq:precode1_fdr_zf_err}
\end{equation}
Then, the SINR's at the MS's and RS are computed as
\begin{eqnarray}
\textrm{SINR}_{1,\gamma_m} &=& \frac{\left|\mathbf{h}^{\textrm{BS}}_{1,\gamma_m} \hat{\mathbf{w}}_{1,{\gamma}_m}^{\textrm{BS}}\right|^2{P}_{1,\gamma_m}}{1+I^{(1)}_{1,\gamma_m}+I^{(2)}_{1,\gamma_m}}, \nonumber \\
\textrm{SINR}_{\textrm{RS}} &=& \frac{ \left| \mathbf{h}^{\textrm{BS}}_{\textrm{RS}} \hat{\mathbf{w}}_{\textrm{RS}}^{\textrm{BS}}\right|^2 {P}_{\textrm{RS}}} {1+I^{(1)}_{RS}+ I^{(2)}_{RS}}, \nonumber \\
\textrm{SINR}_{2,j} &=& \left|{h}_{2,j}^{\textrm{RS}}\hat{w}_{2,j}^{\textrm{RS}} \right|^2 {P}_{2,j},
\label{eq:afterzfbf_fdr_zf_err}
\end{eqnarray}
where 
\begin{eqnarray}
I^{(1)}_{1,\gamma_m} &=& \sum_{k \neq m} \left|\mathbf{h}^{\textrm{BS}}_{1,\gamma_m} \hat{\mathbf{w}}_{1,{\gamma}_k}^{\textrm{BS}}\right|^2{P}_{1,\gamma_k}+\left|\mathbf{h}^{\textrm{BS}}_{1,\gamma_m}\hat{\mathbf{w}}_{\textrm{RS}}^{\textrm{BS}}\right|^2{P}_{\textrm{RS}}, \nonumber \\
I^{(2)}_{1,\gamma_m} &=& \left| \mathbf{h}^{\textrm{BS}}_{1,\gamma_m} \hat{\mathbf{w}}_{2,j}^{\textrm{BS}} + {h}_{1,\gamma_m}^{\textrm{RS}}\hat{w}_{2,j}^{\textrm{RS}} \right|^2 {P}_{\textrm{RS}}, \nonumber \\
I^{(1)}_{RS} &=& \sum_{k} \left| \mathbf{h}^{\textrm{BS}}_{\textrm{RS}} \hat{\mathbf{w}}_{1,{\gamma}_k}^{\textrm{BS}} \right|^2 {P}_{1,\gamma_k}, \nonumber \\
I^{(2)}_{RS} &=& \left|\mathbf{h}^{\textrm{BS}}_{\textrm{RS}}\hat{\mathbf{w}}_{2,j}^{\textrm{BS}}+{h}_{\textrm{RS}}^{\textrm{RS}}\hat{w}_{2,j}^{\textrm{RS}} \right|^2 {P}_{\textrm{RS}}.
\label{eq:afterzfbf_interference}
\end{eqnarray}
Note that we have assumed that ${P}_{\textrm{RS}} = {P}_{2,j}$ in (\ref{eq:afterzfbf_fdr_zf2}), since the incoming information rate and outgoing information rate are the same at the RS. However, it is obvious that $\log_2(1+\textrm{SINR}_{\textrm{RS}}) \neq \log_2 (1+\textrm{SINR}_{2,j})$ with CSI feedback errors, even though the actual transmit power satisfies ${P}_{\textrm{RS}} = {P}_{2,j}$. Therefore, in this case, it may be reasonable to modify the sum rate in (\ref{eq:cap_fdr_zf}) for the FDR with ZFBF as
\begin{eqnarray}
\hat{R}_{\textrm{FDR}} &=& \max_{\begin{smallmatrix} \Gamma \subset \{1, 2, \cdots , N_{1} \} \\ j \in {1, 2, \cdots, N_2} \end{smallmatrix}} \sum_{m=1}^{|\Gamma|}\log_{2} \left( 1+\textrm{SINR}_{1,\gamma_m} \right)\nonumber\\ &+& \min \left\{ \log_2(1+\textrm{SINR}_{\textrm{RS}}), \log_2 (1+\textrm{SINR}_{2,j}) \right\}, \label{eq:cap_fdr_zf_err}
\end{eqnarray}
For the HDR with ZFBF, $\hat{R}_{\textrm{HDR}}$ can also be computed with the CSI feedback errors similarly to the procedure described above.

\section{Numerical Results} \label{sec5}

\begin{figure}[!t]
\centering
\includegraphics[height=2.5in]{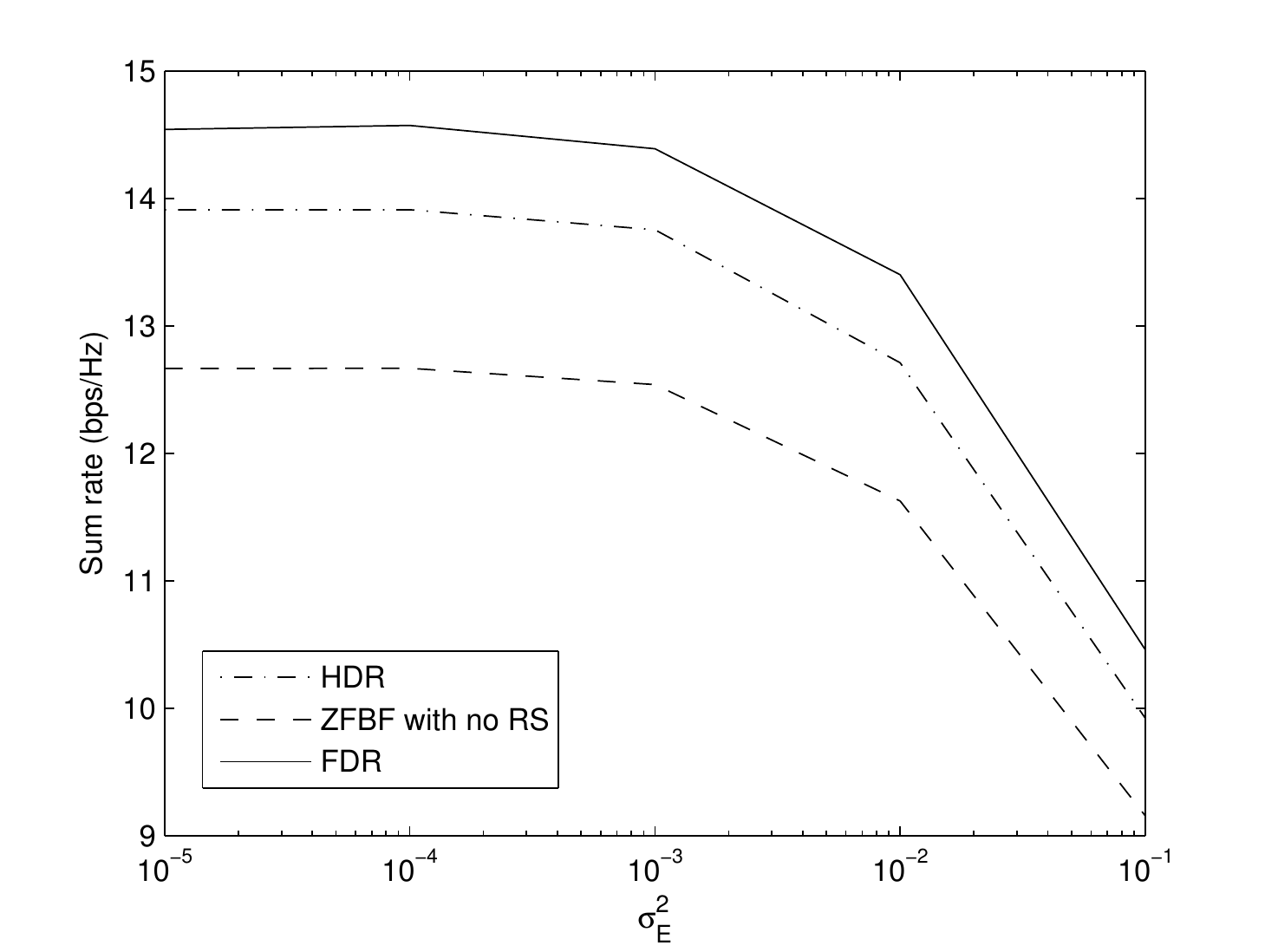}
\caption{Sum rate of relay transmission schemes vs. $\sigma_{E}^{2}$ ($L=2$, $N=4$, $G=20$, $I=10$ and $Q=6$).} \label{fig2}
\end{figure}

\begin{figure}[!t]
\centering
\includegraphics[height=2.5in]{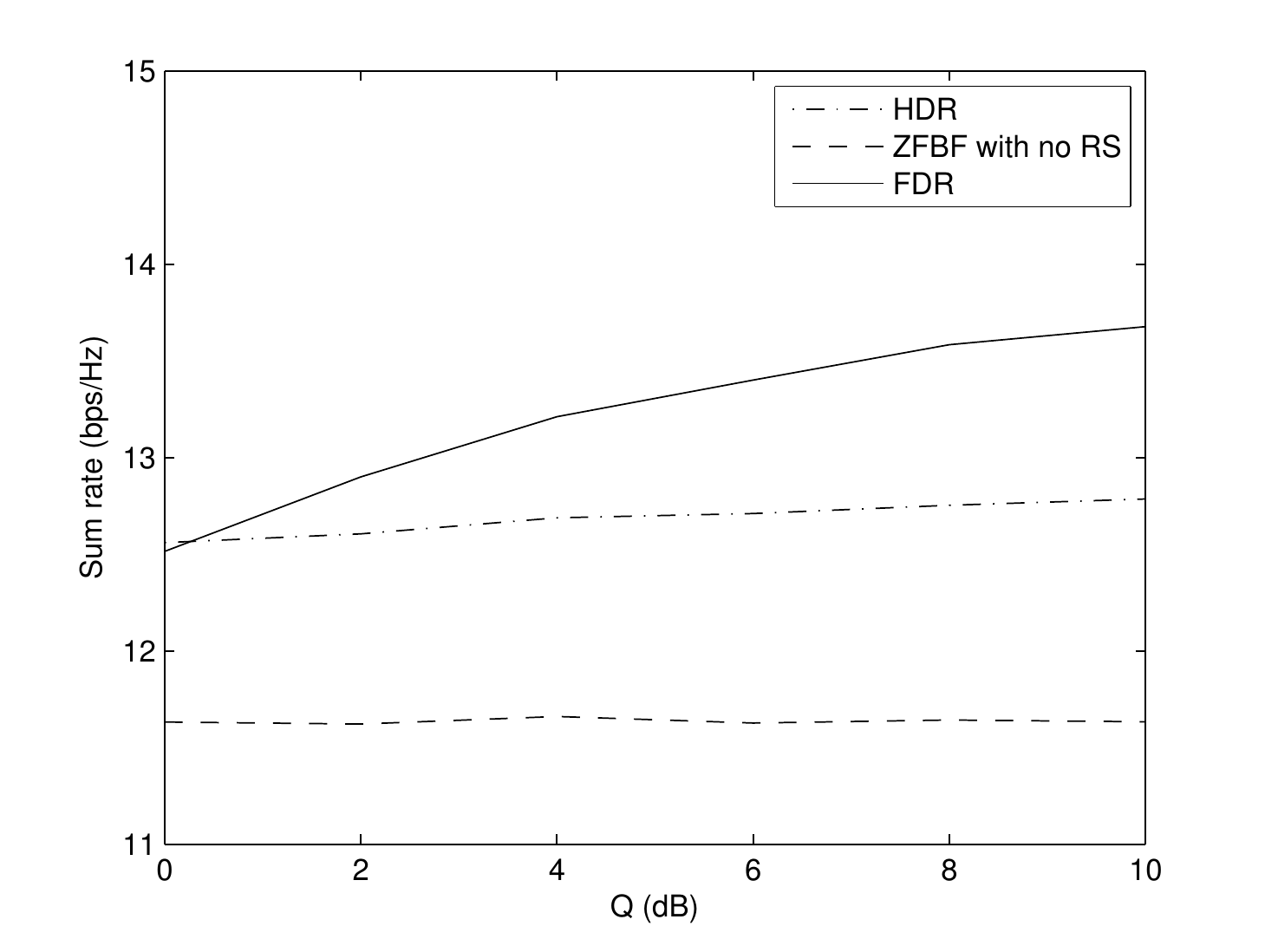}
\caption{Sum rate of relay transmission schemes vs. $Q$ ($L=2$, $N=4$, $G=20$, $I=10$ and $\sigma_{E}^{2}=0.01$).} \label{fig3}
\end{figure}

\begin{figure}[!t]
\centering
\includegraphics[height=2.5in]{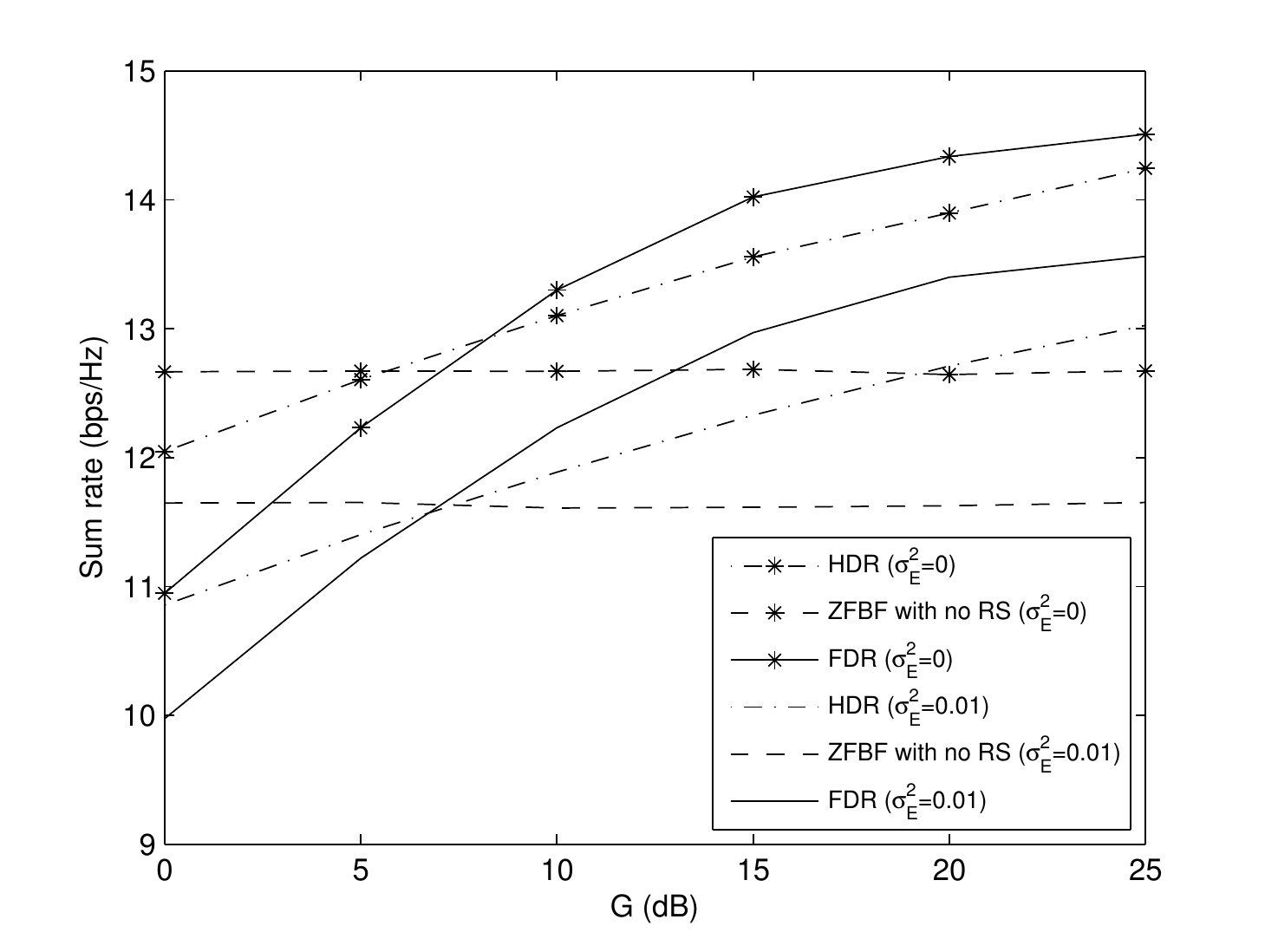}
\caption{Sum rate of relay transmission schemes vs. $G$ ($L=2$, $N=4$, $I=10$, and $Q=6$).} \label{fig4}
\end{figure}

\begin{figure}[!t]
\centering
\includegraphics[height=2.5in]{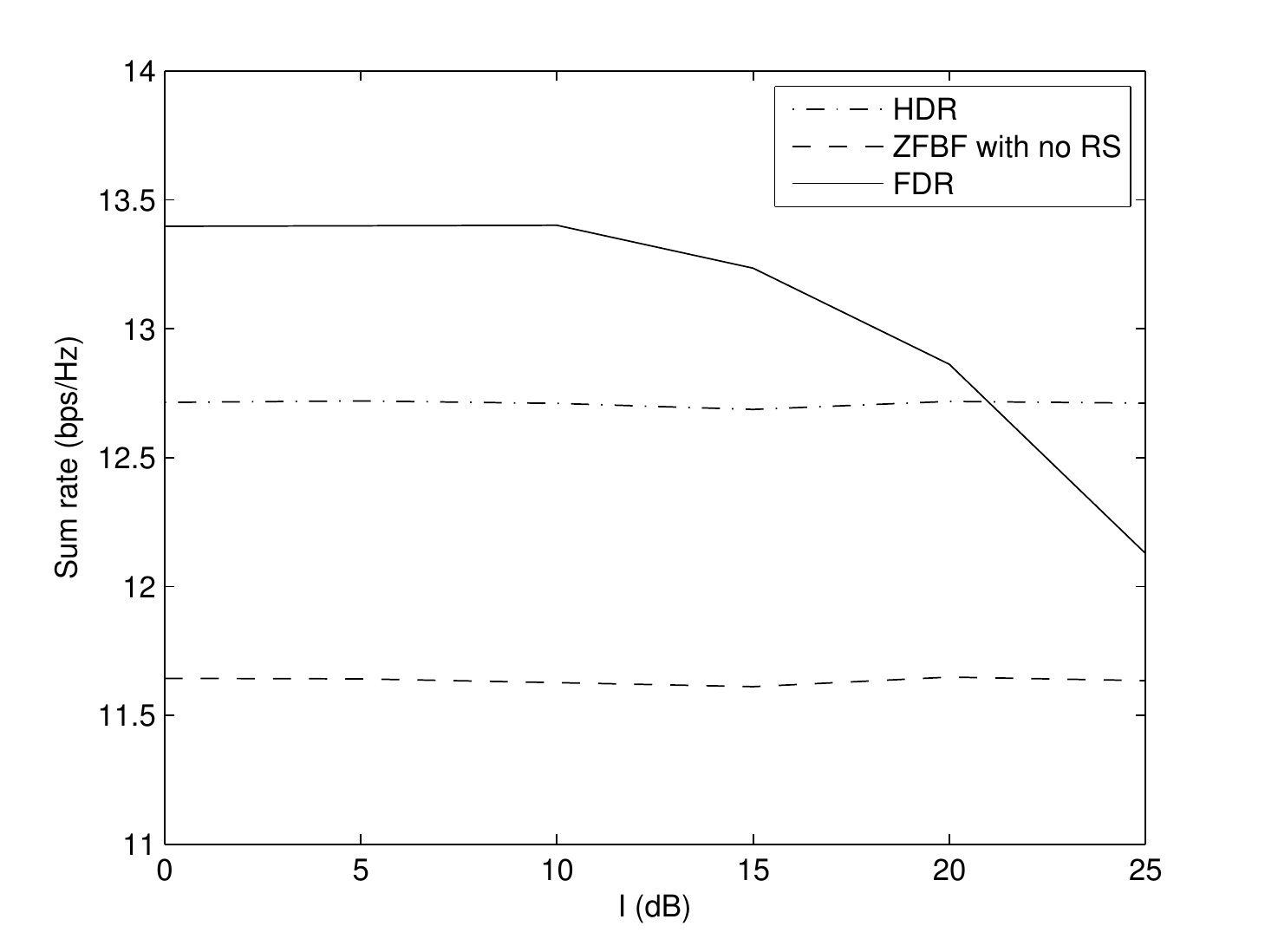}
\caption{Sum rate of relay transmission schemes vs. $I$ ($L=2$, $N=4$, $G=20$, $Q=6$ and $\sigma_{E}^{2}=0.01$).} \label{fig5}
\end{figure}

We evaluate the ergodic sum rate performance of the HDR and FDR schemes through Monte Carlo simulations; 10,000 independent channel realizations are used to obtain the ergodic sum rate of each scheme. We also present the performance of an ordinary multiuser MIMO system without RS \cite{Yoo05} as a baseline. We assume ${P}_{T}^{\textrm{BS}} = 100$ and ${P}_{T}^{\textrm{RS}} = 50$. Moreover, for every MS in the first or second group, channel coefficient of the BS-to-MS link is assumed to follow i.i.d. complex Gaussian distribution with zero mean and unit variance. Similarly, channel coefficient of the RS-to-MS link is assumed to follow the same distribution with zero mean and the variance of $Q$ dB. These assumptions imply that the signal-to-noise ratio (SNR) of the BS-to-MS link and RS-to-MS link is set to 20dB and $(Q+17)$dB, respectively. It is also assumed that the average channel gain of each entry in $\mathbf{h}_{\textrm{RS}}^{\textrm{BS}}$ and that of ${h}_{\textrm{RS}}^{\textrm{RS}}$ are, respectively, $G$-dB and $I$-dB larger than that of the BS-to-MS link. Assuming that the number of MS's in each group is the same, we denote the number of MS's in each group as $N$ (i.e., $N_i = N$ with $i=1,2$). 

In Fig. \ref{fig2}, we presnet the variation of ${\mathcal{E}}\left\{\hat{R}_{\textrm{HDR}}\right\}$ and ${\mathcal{E}}\left\{\hat{R}_{\textrm{FDR}}\right\}$ with $\sigma_{E}^{2}$, when $L=2$, $N=4$, $G=20$, $I=10$, and $Q=6$. The ergodic sum rate of ZFBF without RS \cite{Yoo05} is also presented. The sum rate performance for all transmission schemes is shown to decrease steeply when $\sigma_{E}^{2}>10^{-3}$. Moreover, it is seen that the proposed FDR with ZFBF outperforms the HDR with ZFBF and the ordinary ZFBF without RS, even though the CSI's at the BS are erroneous. 

Fig. \ref{fig3} shows the variation of the ergodic sum rate performance with $Q$, when $L=2$, $N=4$, $G=20$, $I=10$, and $\sigma_{E}^{2}=0.01$. As shown in Fig. \ref{fig3}, even though the RS-to-MS link becomes more reliable, the HDR sum rate is almost constant due to the time sharing. However, the sum rate of the proposed FDR increases steeply compared with the HDR. Fig. \ref{fig4} presents how the ergodic sum rate performance varies with $G$, when $L=2$, $N=4$, $I=10$, and $Q=6$. In Fig. \ref{fig4}, $\sigma_{E}^{2}=0$ indicates that the perfect CSI's are available at the BS. Note that the larger value of $G$ makes the BS-to-RS link more reliable. The FDR with ZFBF is shown to outperform the HDR when $G>8$. 

Fig. \ref{fig5} shows the variation of the ergodic sum rates with $I$, when $L=2$, $N=4$, $G=20$, $Q=6$, and $\sigma_{E}^{2}=0.01$. Note that the larger $I$ causes the stronger self-interference to the RS receiver in the case of the FDR with ZFBF. The HDR scheme must not depend on $I$. For the FDR to outperform the HDR, $I$ must be less than 21. This suggests that sufficient isolation between the transmit and receive antennas at the RS is necessary for the proposed FDR.

\section{Conclusion} \label{sec6}
We have proposed an FDR transmission scheme for the downlink of a cellular system. The FDR has been realized through the use of a ZFBF scheme based on multiple antennas at the base station. Numerical results have verified that the proposed FDR provides the performance improvement over the HDR when the antenna isolation is sufficient for the proposed FDR to operate within a tolerable self-interference range. We have also investigated the impact of CSI errors on the sum rate performance.


\end{document}